\documentclass[english,british]{article}
\usepackage[T1]{fontenc}
\usepackage[utf8]{inputenc}
\usepackage{verbatim}
\usepackage{amsbsy}
\usepackage{amstext}
\usepackage{amssymb}
\usepackage{graphicx}

\makeatletter

\providecommand{\tabularnewline}{\\}

\newcommand{\lyxaddress}[1]{
	\par {\raggedright #1
	\vspace{1.4em}
	\noindent\par}
}

\usepackage{a4wide}
\renewcommand\[{\begin{equation}}
\renewcommand\]{\end{equation}}
\date{}
\usepackage{simplewick}
\usepackage{hyperref,doi}
\usepackage[
        backend=biber,
        style=phys,  
        mcite=true,
        subentry,
        hyperref=true,
        ]{biblatex}
\usepackage[version=4]{mhchem}
\usepackage{tipa}
\newcommand{\lambdabar}{\mbox{\textipa{\textcrlambda}}}

\ifdefined\showcaptionsetup
 \PassOptionsToPackage{caption=false}{subfig}
\fi
\usepackage{subfig}
\makeatother

\usepackage{babel}
\begin{document}
\title{Does chemistry need more physics ?}
\author{Trond Saue}
\maketitle

\lyxaddress{\begin{center}
Laboratoire de Chimie et Physique Quantique,\\
UMR 5626 CNRS - Université Toulouse III-Paul Sabatier,\\
 118 Route de Narbonne, F-31062 Toulouse, France\\
mail: trond.saue@irsamc.ups-tlse.fr
\par\end{center}}
\begin{abstract}
In this mini-review I look into the physics underlying the theory
of electronic structure of atoms and molecules. Quantum mechanics
is needed to understand the structure of the periodic table. Special
relativity is indispensable for a correct description of the chemistry
of the heavy elements. With increased accuracy of quantum chemical
calculations, it is natural to ask if chemistry needs more physics. 
\end{abstract}

\section{Quantum mechanics}

The periodic table sets the stage for all of chemistry. Its date of
birth is usually considered to be February 17, 1869, when Mendeleev
sent the first compilation to print.\cite{Hargittai_SC2019} An understanding
of the periodicity of chemical and physical properties observed when
ordering the elements known at the time according to atomic mass required
elucidation of the structure of atoms, and here quantum mechanics
plays a major role.\cite{Pyykko_CR2012,Schwerdtfeger_NRC2020} In
this respect it is interesting to note that the idea of atoms put
forward by Democritus represents the quantisation of matter in indivisible
parts. A decisive step was the gold foil experiment of Geiger and
Marsden \cite{Geiger_Marsden_PRS1909} which led their mentor Rutherford
to propose a model of the atom with \textsl{`a charge $\pm Ne$ at
its centre surrounded by a sphere of electrification containing a
charge $\mp Ne$ supposed uniformly distributed throughout a sphere
of radius $R$}'.\cite{Rutherford_1911} The scattering experiment
did not allow to determine the sign of the central charge, but in
a subsequent paper, Rutherford spells out his choice and also provides
a name for the central charge:\cite{Rutherford_PM1912} \textsl{In
a previous paper I have given reasons for believing that the atom
consists of a positively charged nucleus of very small dimensions,
surrounded by a distribution of electrons in rapid motion, possibly
of rings of electrons rotating in one plane. }Moseley was able to
determine the nuclear charge of some 40 elements by X-ray emission
spectroscopy and made it clear that the position of an element in
the periodic table is determined by its atomic number rather than
atomic mass,\cite{Moseley_PM1913,Moseley_PM1914} as already proposed
by van der Broek.\cite{vdBroek_PZ1913} Rutherford also pointed out
that the gold foil experiment was incompatible with the `plum pudding'
model proposed by Thomson,\cite{Thomson_LEDPMJS1904} in which `\textsl{the
atoms of the elements consist of a number of negatively electrified
corpuscles enclosed in a sphere of uniform positive electrification}'.
Corpuscule was the name given by J. J. Thomson to the subatomic \emph{particle}
whose discovery he reported in 1897.\cite{Thomson_PM1897} The name
electron was originally proposed by Stoney as the atom of electricity,
\cite{Stoney1894,OHara} that is, a \emph{quantisation} of charge.
J. J. Thomson kept using the term \emph{corpuscule}, for instance
in his 1906 Nobel lecture,\cite{Thomson_Nobel1906} a model of clarity,
and only gave in to the term \emph{electron} in his 1914 book on atomic
theory.\cite{thomson_atomic_theory_1914} His son G.P. Thomson was
awarded the 1937 Nobel prize in physics for the discovery of electron
diffraction, which led Max Born to comment\cite{Born_thomson-obituary}
\begin{quote}
\textsl{It is a fascinating fact that father and son have given the
most striking evidence for the apparently contradictory properties
of the electron: the father proving its character as a particle, the
son its character as a wave. Modern quantum theory has shown that
these are two aspects of the same phenomenon, depending on different
kinds of observation: not contradictory, but complementary.}
\end{quote}
A fundamental problem with the planetary model of the atom proposed
by Rutherford (see also Ref. \cite{Perrin_RS1901}) was that is was
inherently unstable. According to classical electrodynamics the circulating
electrons would radiate, loose energy and spiral into the nucleus.
About this Rutherford simply stated: `T\textsl{he question of the
stability of the atom proposed need not be considered at this stage,
for this will obviously depend upon the minute structure of the atom,
and on the motion of the consitutent charged parts.}'\cite{Rutherford_1911}
This problem was solved by Bohr who postulated the existence of stationary
states of electrons associated with the quantisation of angular momentum.\cite{Bohr_PM1913_I}
Bohr furthermore proposed that emission and absorption of light was
associated with transitions between these stationary states. His model
thereby not only explained the stability of atoms, but also the discrete
nature of their spectra.

The elaboration of his model led Bohr to ponder on atomic structure,
notably `\textsl{to find configurations and motions of the electrons
which would seem to offer an interpretation of the variations of the
chemical properties of the elements with the atomic number as they
are so clearly exhibited in the well-known periodic law}.'\cite{Bohr_Nature1921a}
Bohr cites the work of Kossel and Lewis. The former stressed the stability
of closed shells.\cite{Kossel_AP1916} In the same year (1916), Lewis
published his model of the cubical atom,\cite{Lewis_JACS1916} where
appears in a natural manner shell structure and the separation of
core and valence, as well as the covalent bond. Interestingly, his
notes on the subject goes back to 1902.\cite{Kohler_HSPS1971} An
important contribution from Bohr was the \emph{Aufbau} principle,\cite{Bohr_ZP1923}
which Mulliken describes in the following manner:\cite{Mulliken_PhysRev.32.186,Kragh_Qatom2012}\textsl{Bohr's
method of determining electron configuration in atoms by imagining
all electrons removed, and then returning them one by one, each to
the available orbit of lowest energy.}' However, further theory was
needed in order to justify why electrons did not pile up in the same
orbits. This came in the form of Pauli's exclusion principle,\cite{Pauli_ZP1925}
which is a particular instance of the spin-statistics theorem.\cite{Curceanu_AJP2012}
In the same paper Pauli introduced the fourth quantum number $m_{s}=\pm1/2$,
later to be associated with spin. 

Despite initial successes, the Bohr atom model accumulated failures.\cite{Margenau_AJP1944,Kragh_Qatom2012}
For chemists, an immediate concern was that it was unable to describe
bonding in the \ce{H2} molecule. Furthermore, it predicted an ionisation
energy of helium of 28.9 eV, which is more than 4 eV too large.\cite{Hylleraas_AQC1964}
Scerri writes:\cite{Scerri_PT2} `\textsl{Only following the advent
of quantum mechanics, as distinct from the old quantum theory, was
there a possibility of calculating the energy of helium, and even
then only approximately.}' There may be differing opinion as to what
is exact, but we may note that the energy suggested by Bohr was -3.605
$E_{h}$. In 1928 Hylleraas, using the new quantum mechanics, obtained
the value -2.895 $E_{h}$\cite{Hylleraas_ZfP1928} which he subsequently
improved to -2.9037 $E_{h}$ the year after.\cite{Hylleraas_ZfP1929}
In 2007 Nakashima and Nakatsuji reported the calculation of the non-relativistic
ground state energy of the helium atom with 40-digit accuracy,\cite{Nakatsuji_JCP2007}
and the five first digits coincide with the 1929 Hylleraas value.

A much more detailed account of the development of the periodic table
and our understanding of it has been provided by Scerri,\cite{Scerri_PT2}
but he expresses doubts as to what extent quantum mechanics can explain
the periodic system. We read for instance: \textsl{Theoretical calculations
cannot actually predict the electronic configuration for any element.}
Let us consider this proposition in some detail. First we note that
it assumes that atomic electronic structure reduces to the specification
of an electron configuration. The IUPAC definition of an electronic
configuration is\cite{IUPAC_2019goldbook}
\begin{quote}
\textsl{A distribution of the electrons of an atom or a molecular
entity over a set of one-electron wavefunctions called orbitals, according
to the Pauli principle. From one configuration several states with
different multiplicities may result.}
\end{quote}
The definition is clearly formulated in the language of quantum mechanics.
Calculations of the electronic structure of atoms (and molecules),
within the Born--Oppenheimer approximation, are based on an electronic
Hamiltonian of the generic form
\begin{equation}
\hat{H}={\displaystyle \sum_{i=1}^{N}\hat{h}(i)+\frac{1}{2}\sum_{i\ne j}^{N}\hat{g}(i,j)+V_{nn},}\label{eq:electronic=000020Hamiltonian}
\end{equation}
where $V_{nn}$ is the classical repulsion of fixed nuclei. In the
absence of external fields, the one-electron Hamiltonian has the general
form
\begin{equation}
\hat{h}=\hat{h}_{0}+V,
\end{equation}
where $V=-e\phi_{\text{nuc}}$ represents the interaction with the
electrostatic potential of clamped nuclei. In non-relativistic calculations
the free-particle operator is simply the operator of kinetic energy

\begin{equation}
\hat{h}_{0}^{NR}=\frac{p^{2}}{2m_{e}},
\end{equation}
whereas the two-electron operator is given by the instantaneous Coulomb
interaction
\begin{equation}
\hat{g}^{C}\left(1,2\right)={\displaystyle \frac{e^{2}}{4\pi\varepsilon_{0}r_{12}}}.\label{eq:g^C}
\end{equation}
It is the two-electron term that bars the exact solution of the electronic
problem. In its absence one would be able to solve it by separation
of variables, and the exact solution would be given by a (Hartree)
product of one-electron wave-functions, denoted orbitals by Mulliken.\cite{Mulliken_PhysRev.41.49,Mulliken_Science1967}
The orbital product has to be antisymmetrized to comply with the Dirac--Fermi
statistics of electrons. The resulting Slater determinant\cite{Slater_PhysRev.34.1293}
is nevertheless taken as a starting point for the Hartree--Fock method,
which provides a mean-field description of the electronic interaction.
With such a wave function electronic structure is reduced to a specification
of what orbitals are occupied. This is in fact more precise than an
electron configuration since the latter only specifies what electronic
\emph{shells} ($s,\,p,\,d,\,f$) are occupied. At this point Scerri
comments:\cite{Scerri_PT2} `\textsl{..the very notion of a particular
number of electrons in a particular shell stands in violation of the
Pauli exclusion principle, which states that electrons can not be
distinguished. The indistinguishability of electrons implies that
one can never state that a particular number of electrons are in any
particular shell,$\ldots$}'. This is, however, a clear misunderstanding;
the anti-symmetrisation of the orbital product means that all electrons
have equal probability of occupying a particular orbital.

As already stated by the IUPAC definition above a configuration can
give rise to several electronic states $^{2S+1}L$, in non-relativistic
notation. State-specific mean-field calculations are therefore more
naturally based on \emph{configuration state functions} (CSFs), which
in the atomic case are fixed linear combinations of Slater determinants
which are eigenfunctions of operators $\hat{L}^{2},\,\hat{L}_{z},\,\hat{S}^{2}$
and $\hat{S}_{z}$.\cite{Boys_RSPA1950.125} Early on it was realized
that accurate calculations required the inclusion of more than one
configuration, leading to the development of the configuration interaction
(CI) method.\cite{Shavitt_MP1998} For instance, in 1950 Boys reported
a CI calculation on the ground state of the beryllium atom using 10
CSFs and a \texttt{3s1p} Slater-type orbital basis.\cite{Boys_RSPA1950.125}
He reports that the \ce{1s^2 2s^2} configuration obtained a coefficient
of 0.892 (after normalization of the CI-wave function to unity), whereas
the second most important CSF was the \ce{1s^2 2p^2} configuration
with coefficient 0.307. The latter is certainly non-negligible, but
translates into a weight of 9.4\%, meaning that the ground-state configuration
\ce{1s^2 2s^2} listed for beryllium is a good approximation. I have
not been able to find a tabulation of the composition in terms of
configurations for all elements of the periodic table, but recently
Shabaev and co-workers reported a CI-study of the ground-state configurations
of superheavy elements (SHEs) with atomic numbers 120 \ensuremath{\leqslant}
Z \ensuremath{\leqslant} 170. They report:'\textsl{We obtain that
in spite of the complex electronic structure of the considered SHEs,
the ground-state levels have distinct dominant configurations with
the weights exceeding 0.85.}'\cite{Savelyev_PhysRevA.107.042803}
It therefore appears that the notion of a dominant ground-state electronic
configuration is a good approximation. However, it remains an approximation,
just as orbitals. An advantage is that it provides a language for
a concise description of the electronic structure of the elements
of the periodic table. In the same manner one may deduce simple rules
for the filling of orbitals, such as the Madelung--Janet $n+\ell$
rule.\cite{Schwerdtfeger_NRC2020} However, one should not forget
that such rules have been deduced from simple models and that any
exception to such rules does not invalidate quantum mechanics.

Finally, before closing this section, I would like to comment on the
relation between theory, experiment and explanation. Rohrlich gives
a thought-provoking statement:\cite{Rohrlich_classical}
\begin{quote}
\textsl{Having gone through all the trouble that is necessary to find
the `right' laws and the `right' theory, no one should be surprised
that the predictions of the theory indeed agree with experiment or,
for that matter, that a given phenomenon can be fully accounted for
by the theory. Nevertheless, we say that the theory }\textit{explains}\textsl{
the phenomenon. In this sense }\textit{scientific explanation}\textsl{
is circular. The emphasis should really be on the existence and correctness
of the deductive system, not on the explanation, for the theory is
}\textit{not derived}\textsl{ or derivable from experiments. It is
a mental step into the abstract and general, postulating validity
in the future and for all experiments including those that have never
been carried out before. The existence of a valid theory is therefore
the nontrivial and indeed the very remarkable facet of scientific
exploration.}
\end{quote}

\section{Relativistic quantum mechanics }
\begin{center}
\begin{figure}[h]
\begin{centering}
\includegraphics[width=0.9\textwidth]{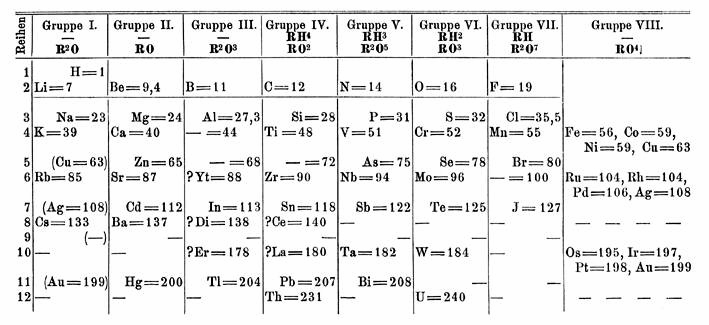}
\par\end{centering}
\caption{Mendeleev's 1871 periodic table \cite{Mendelejeff_1871table} (From
Wikipedia, the free encyclopedia)}\label{fig:Mendeleev1871}
\end{figure}
\par\end{center}

We return to the periodic table. Figure \ref{fig:Mendeleev1871} shows
the 1871 version of the periodic table\cite{Mendelejeff_1871table}.
Mendeleev was not only able to accommodate the chemical elements known
at the time, but his table strongly suggested missing elements, for
which Mendeleev, in view of the periodicity, predicted chemical and
physical properties. Particularly successful were his predictions
for the elements below (Sanskrit \emph{eka}) aluminium, boron and
silicon, discovered 1875, 1879 and 1886 and given the names gallium,
scandium and germanium, respectively, in a particularly nationalistic
period of European history. Interestingly, Mendeleev also made predictions
for elements lighter than hydrogen, clearly with less success.\cite{Scerri_PT2} 

Figure \ref{fig:Mendeleev1871} also suggests missing elements in
the lower part of the table. Here Mendeleevs predictions would be
less  convincing since the periodic trends are broken. The two major
causes for this are i) relativistic effects and ii) the lanthanide
contraction. Whereas Albert Einstein is very well known as the father
of the theory of relativity, it is less known that the term 'lanthanide
contraction' was coined by Viktor Moritz Goldschmidt, the father of
geochemistry.\cite{goldschmidt:lanthanidecontraction} The two actually
met in 1920, when Goldschmidt invited Einstein to Oslo to lecture
about relativity (cf. Figure \ref{fig:Goldschmidt-Einstein})
\begin{center}
\begin{figure}[h]
\begin{centering}
\includegraphics[width=0.7\columnwidth]{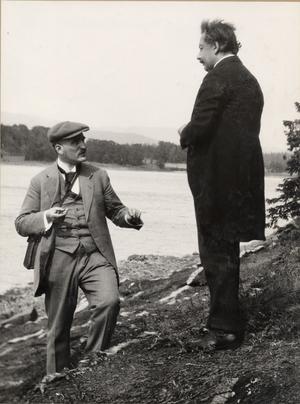}
\par\end{centering}
\caption{Viktor Moritz Goldschmidt and Albert Einstein at Langodden in the
inner Oslofjord in Norway 1920. (Photo: Halvor Rosendahl/Museum of
University History , University of Oslo) }\label{fig:Goldschmidt-Einstein}
\end{figure}
\par\end{center}

\textbf{Relativistic effects} may be defined as the difference between
the actual universe in which the speed of light is finite, fixed at
299 792 458 m/s,\cite{Tiesinga_RevModPhys.93.025010} and a hypothetical
non-relativistic universe, with infinite speed of light. \footnote{It is interesting to note that it was not until the 1676 experiments
of Ole Rømer that it became clear that the speed of light was finite.\cite{Tuinstra_PT2004}} A diagnostic of relativistic effects is provided by the Lorentz factor
\begin{equation}
\gamma=\frac{1}{\sqrt{1-v^{2}/c^{2}}},
\end{equation}
where $v$ is the speed of the object and $c$ the speed of light;
in the non-relativistic limit $c\rightarrow\infty$, the Lorentz factor
tends to one. In atomic and molecular systems relativistic effects
arise from the high speeds of electrons in the vicinity of highly
charged nuclei. For hydrogen-like atoms the average speed of a $1s$
electron is $v_{1s}=Z$ in atomic units, compared to the speed of
light in the same units, $c=137.035999177\,a_{0}E_{h}/\hbar$. For
heavy atoms the Lorentz factor accordingly becomes significantly larger
than one and from the relativistic mass increase $m=\gamma m_{0}$
a contraction of atomic orbitals is expected\cite{Pyykko_AQC1978,Pyykko_ACR1979,Pyykko_CR1988}.
Such a contraction had in fact been reported by Burke and Grant in
1967.\cite{Burke_Grant_PRS1967} In a many-electron atom, however,
there will be a competing effect from the increased screening of the
nuclear charge due to such contraction. In practice one finds that
$s$ and $p$ orbitals contract, whereas $d$ and $f$ orbitals expand. 

Relativistic effects profoundly modify the chemistry of the heavy
elements. Without relativity
\begin{itemize}
\item gold would have the same colour as silver\cite{Pyykko_ACR1979,Pyykko_CR1988,Romaniello_JCP2005}
\item mercury would not be liquid at room temperature\cite{Steenbergen_JPCL2017}
\item your car would not start, at least if it has a lead battery\cite{Pekka_PhysRevLett.106.018301}
\end{itemize}
to mention some iconic examples.

The \textbf{lanthanide contraction} is often viewed as a \emph{horizontal}
effect. It is observed that the radius of trivalent cations of the
$4f$-elements, the lanthanide series, is reduced from 117.2 (\ce{La^3+})
to 100.1 pm (\ce{Lr^3+}). However, this is a perfectly normal trend
along a row in the periodic table; the addition of electrons does
not outweigh the effect of increased nuclear charge. In fact, as pointed
out by Lloyd,\cite{Lloyd_JCE1986} the contraction across these 15
elements is less than than across the 11 third-row elements Ca$^{2+}$
- Zn$^{2+}$ (114 - 88 pm), filling in $3d$ electrons. The lanthanide
contraction should rather be viewed as a \emph{vertical} effect, as
seen for the noble metals. When going from copper to silver, the covalent
radius increases (from 138 pm to 153 pm), as expected when going down
a column in the periodic system. However, when proceeding to gold,
the covalent radius is reduced to 144 pm. An explanation for this
is that going from copper to silver implies adding complete $s$-
, $p$- and $d$-shells, whereas when continuing to gold one adds
$s,p,d$-shells as well as a complete $f$-shell, with increasingly
reduced screening of the nuclear charge.
\begin{center}
\begin{figure}
\begin{centering}
\includegraphics[width=0.9\textwidth]{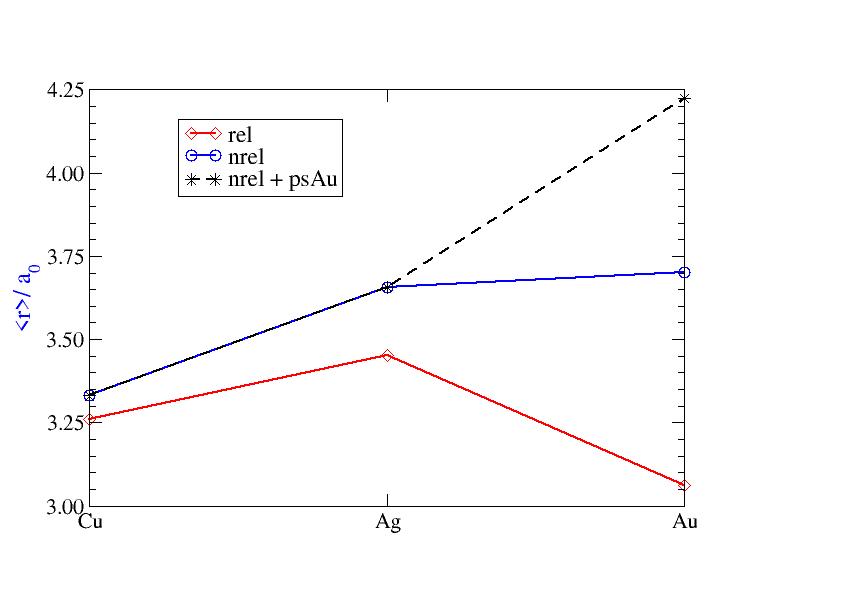}
\par\end{centering}
\caption{Radial expectation values $\left\langle r\right\rangle _{ns}$ (in
$a_{0}$) of valence $ns$ orbitals of noble metals, calculated at
the Hartree--Fock level. Data from Ref.\cite{Bagus_CPL1975}.}\label{fig:bagus1975}

\end{figure}
\par\end{center}

Neither relativistic effects nor the lanthanide contraction can be
directly observed by experiment. Instead, these effects can be explored
by theoretical chemists. An elegant example of this is provided by
a 1975 Hartree--Fock study by Bagus and co-workers.\cite{Bagus_CPL1975}
As a measure for the covalent radius they calculated the radial expectation
values $\left\langle r\right\rangle _{ns}$ of the valence $ns$ orbitals
of the noble metals, as seen in Figure \ref{fig:bagus1975}. A relativistic
calculation reproduces the anomalous evolution of radii, whereas a
non-relativistic calculation makes gold slightly bigger than silver.
Bagus \emph{et al.} furthermore attempted to model the lanthanide
contraction. We have seen that it originates from the filling of the
$4f$-shell. A simple solution was therefore to not include the $4f$-shell
in the calculation. This would lead to a surplus charge of +14 of
the atom, so the nuclear charge was reduced accordingly, formally
corresponding to an excited state of terbium. With this pseudo-gold
atom, an almost linear increase of the covalent radii is observed
down the noble metal series. This model has also been extended to
molecules.\cite{Fossgaard_JCP2003} It should be remarked, though,
that the relativistic effects and lanthanide contraction are not additive.
\begin{center}
\begin{figure}
\begin{centering}
\includegraphics[width=0.75\textwidth]{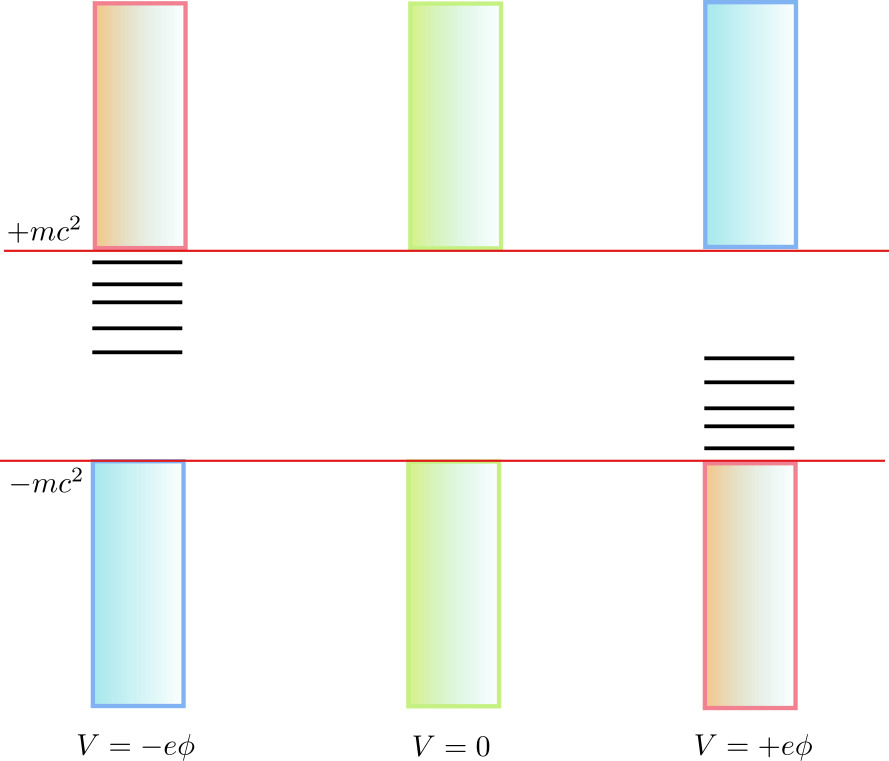}
\par\end{centering}
\caption{Spectrum of the Dirac equation.}\label{fig:Spectrum-of-the}

\end{figure}
\par\end{center}

Let us now investigate what relativity brings to quantum mechanics,
and \emph{vice versa}. The starting point for conventional relativistic
atomic and molecular calculations is an electronic Hamiltonian of
the form Eq.\textasciitilde (\ref{eq:electronic=000020Hamiltonian})
where the free-particle one-electron operator is given by the Dirac
Hamiltonian
\begin{equation}
\hat{h}_{0}^{R}=\beta m_{e}c^{2}+c\left(\boldsymbol{\alpha}\cdot\boldsymbol{p}\right).\label{eq:Dirac=000020Hamiltonian}
\end{equation}
In the free-particle case ($V=0$) the energy spectrum, shown in Figure
\ref{fig:Spectrum-of-the}, agrees with the energy expression 
\begin{equation}
E_{0}=\pm\sqrt{m^{2}c^{4}+c^{2}p^{2}}
\end{equation}
 of relativistic \emph{classical} mechanics, featuring continua of
positive and negative energy, separated by a gap of $2m_{e}c^{2}\approx1.0$
MeV. However, there is a fundamental difference, as pointed out by
Dirac:\cite{Dirac:1934:TDP} in classical mechanics energy is a \emph{continuous}
variable, and so the troublesome negative-energy solutions can be
discarded as being unphysical. This is not the case for quantum mechanics,
since it allows quantum leaps. For atoms (and molecules) the situation
becomes catastrophic: any bound electron will have a finite probability
of a transition into the negative-energy continuum. Matter would not
be stable, even at the scale of nanoseconds,\cite{Tamm_ZP1930,Oppenheimer_PhysRev.35.562}
and as the electron continues down the negative-energy branch, it
will liberate an infinite amount of energy. To prevent this, Dirac
suggested that all (or most) negative-energy solutions are occupied,
in which case the Pauli exclusion principle blocks further occupation.
Dirac further argued that the resulting Dirac sea is not observable,
being uniform over all space, in the absence of other charges. On
the other hand, with sufficient energy, an electron, of charge $-e$,
can be excited out the Dirac sea, leaving a hole, which, from conservation
of charge, would be of charge $+e$. After first falsely suggesting
that this particle would be a proton,\cite{Dirac_PRSA1930} Dirac
in 1931 suggested the existence of an hitherto unknown particle, the
positron. \cite{Dirac_RSPA1931} Incredibly, this anti-particle of
the electron was observed in cosmic radiation within the next year.\cite{Anderson_PhysRev.43.491}

A further curious feature of the Dirac Hamiltonian is the velocity
operator obtained by the Heisenberg equation of motion
\begin{equation}
\hat{\boldsymbol{v}}=\frac{d\mathbf{r}}{dt}=\left\{ \begin{array}{lcl}
{\displaystyle \frac{1}{i\hbar}}\left[\mathbf{r},\hat{h}^{\textrm{NR}}\right] & = & {\displaystyle \frac{\hat{\boldsymbol{p}}}{m_{e}}}\\
\\{\displaystyle \frac{1}{i\hbar}}\left[\mathbf{r},\hat{h}^{\textrm{R}}\right] & = & c\boldsymbol{\alpha}
\end{array}\right..\label{eq:velocity=000020operator}
\end{equation}
 In the non-relativistic case, we find $\hat{\boldsymbol{v}}^{\text{NR}}=\hat{\boldsymbol{p}}/m_{e}$,
which is consistent with the definition of linear momentum from classical
mechanics. In the relativistic case, however, we end up with the peculiar
operator $\hat{\boldsymbol{v}}^{\text{R}}=c\boldsymbol{\alpha}$,
as first pointed out by Breit.\cite{Breit_PNAS1928} Since the Dirac
matrices satisfy~$\alpha_{x}^{2}=\alpha_{y}^{2}=\alpha_{z}^{2}=I_{4}$,
their eigenvalues are of order unity, which in turn implies that each
component of the relativistic velocity operator is on the order of
the speed of light, suggesting superluminal speed of the electron,
in contradiction with the speed limit $c$ imposed on any signal by
relativistic theory. Fortunately, quantum mechanics comes to the rescue:
The Dirac $\boldsymbol{\alpha}$ matrices do not commute, and we can
therefore not measure simultaneously individual components of velocity.\cite{Moss}
The form of the relativistic velocity operator is associated with
\emph{Zitterbewegung}, an oscillatory motion of the electron superimposed
on its mean velocity.\cite{Schrodinger_SPAW1930} One interpretation
this trembling motion is that in the vicinity of an electron its field
is sufficiently strong to allow the creation of an electron-positron
pair. The positron annihilates the original electron and the ``new''
electron takes over.\cite{Saue_CPC2011} This would be imply that
the Dirac equation describes a relay of electrons instead of a single
one ! In face of this dazzling perspective, it is appropriate to cite
Mattuck\cite{mattuck_guide_1992}
\begin{quote}
\textsl{How many bodies are required before we have a problem ? G.
E. Brown points out that this can be answered by a look at history.
In eighteenth-century Newtonian mechanics, the tree-body proble was
insoluble. With the birth of relativity around 1910 and quantum electrodynamics
in 1930, the two- and one-body problems became insoluble. And within
modern quantum field theory, the problem of zero bodies (vacuum) is
insoluble. So, if we are out after exact solutions, no bodies at all
is already too many !}
\end{quote}
Free-particle solutions are of moderate interest in chemistry. The
construction of the electronic Hamiltonian, Eq. (\ref{eq:electronic=000020Hamiltonian}),
proceeds via the introduction of the fields or, in practice, the electromagnetic
potentials of other charges in the system, using the recipe
\begin{eqnarray}
\boldsymbol{p} & \rightarrow & \boldsymbol{p}-q\boldsymbol{A}\\
E_{0} & \rightarrow & E_{0}-q\phi,
\end{eqnarray}
known as the principle of minimal electromagnetic coupling.\cite{Gell-Mann_NC1956}
The coupling is minimal in the sense that only one intrinsic particle
property needs to be specified, namely particle charge $q$. The above
minimal couplings can be written in a more compact form 
\begin{equation}
p^{\mu}\rightarrow p^{\mu}-qA^{\mu}
\end{equation}
using 4-momentum $p^{\mu}=\left(E_{0}/c,\boldsymbol{p}\right)$ and
4-potential $A^{\mu}=\left(\phi/c,\boldsymbol{A}\right)$. This manifestly
Lorentz-invariant form clearly points to the relativistic nature of
the coupling of particles to fields. Interestingly, the minimal substitutions
are a consequence of the interaction Lagrangian proposed by Schwarzschild
in 1903,\cite{Schwarzschild_GN1903} two years before the \emph{annus
mirabilis} of Einstein.

Using minimal electromagnetic coupling, the Dirac Hamiltonian expands
into
\begin{equation}
\hat{h}^{R}=\beta m_{e}c^{2}+c\left(\boldsymbol{\alpha}\cdot\boldsymbol{p}\right)+ec\left(\boldsymbol{\alpha}\cdot\boldsymbol{A}\right)-e\phi.\label{eq:Rham}
\end{equation}
Here we set particle charge equal to electron charge, $q=-e$, which
implies that all solutions, of both positive and negative energy,
are \emph{electronic}. In the atomic case, the spectrum is shown schematically
in the left panel of Figure \ref{fig:Spectrum-of-the}. Corresponding
positronic solutions are obtained by setting $q=+e$ and are shown
in the rightmost panel of Figure \ref{fig:Spectrum-of-the}. Simple
visual inspection shows that the electronic and positronic solutions
are related; the positronic spectrum is simply the electronic one
turned upside down, whereas corresponding eigenfunctions are connected
in a more complicated manner. The underlying symmetry is \emph{charge
conjugation}.\cite{Kramers_1937}

Minimal substitution can also be used to construct the two-electron
interaction. This was the starting point for Charles Galton Darwin
in a 1920 paper which predates quantum mechanics.\cite{Darwin_PhilMag1920}
Writing the interaction of two moving charges as
\begin{equation}
E_{int}=q_{1}\phi_{2}-q_{1}\boldsymbol{v}_{1}\boldsymbol{A}_{2},
\end{equation}
he would next have to plug in the scalar and vector potential of charge
$q_{2}$. In the relativistic Lorentz gauge the general expression
for the potentials are\cite{jackson,nrsbook}
\begin{equation}
\phi(\mathbf{r}_{1},t)=\frac{1}{4\pi\varepsilon_{0}}\int\frac{\rho(\mathbf{r}_{2},t_{r})}{\left|\mathbf{r}_{1}-\mathbf{r}_{2}\right|}d^{3}\boldsymbol{r}_{2};\quad\mathbf{A}(\mathbf{r}_{1},t)=\frac{1}{4\pi\varepsilon_{0}c^{2}}\int\frac{\mathbf{j}(\mathbf{r}_{2},t_{r})}{\left|\mathbf{r}_{1}-\mathbf{r}_{2}\right|}d^{3}\boldsymbol{r}_{2}\label{eq:Lorentz=000020gauge=000020potentials}
\end{equation}
where appears retarded time $t_{r}=t-r_{12}/c$ associated with the
finite speed of light between source and observer, a phenomenon well
known to anyone who has looked at the starry sky; we then look not
only into space, but also back in time. In the special case of point
charges, the above expressions reduce to the Liénard--Wiechert potentials.
Darwin carried out a transformation to Coulomb gauge, without explicitly
stating so. The general expressions are well known\cite{jackson,nrsbook}
\begin{equation}
\phi(\mathbf{r}_{1},t)=\frac{1}{4\pi\varepsilon_{0}}\int\frac{\rho(\mathbf{r}_{2},t)}{\left|\mathbf{r}_{1}-\mathbf{r}_{2}\right|}d^{3}\boldsymbol{r}_{2};\quad\mathbf{A}(\mathbf{r}_{1},t)=\frac{1}{4\pi\varepsilon_{0}c^{2}}\int\frac{\mathbf{j}_{\perp}(\mathbf{r}_{2},t_{r})}{\left|\mathbf{r}_{1}-\mathbf{r}_{2}\right|}d^{3}\boldsymbol{r}_{2},\label{eq:Coulomb=000020gauge=000020potentials}
\end{equation}
and notably features the instantaneous Coulomb interaction (no retarded
time), carried by the scalar potential, whereas the vector potential
is limited to the divergence-free part of the current density, denoted
$\mathbf{j}_{\perp}$. Darwin, however, gives a truncated interaction,
correct only to order $c^{-2}$
\begin{equation}
E_{int}\sim\frac{1}{4\pi\varepsilon_{0}}\left(\frac{q_{1}q_{2}}{r_{12}}-\frac{q_{1}\mathbf{v}_{1}\cdot q_{2}\mathbf{v}_{2}}{2c^{2}r_{12}}-\frac{\left(q_{1}\mathbf{v}_{1}\cdot\mathbf{r}_{12}\right)\left(q_{2}\mathbf{v}_{2}\cdot\mathbf{r}_{12}\right)}{2c^{2}r_{12}^{3}}\right).\label{eq:Darwin=000020interaction}
\end{equation}
One argument that he gives is that the Larmor formula associated with
the radiation of an accelerated electron features the speed of light
to the inverse \emph{cubic} power ($c^{-3}$) and therefore introduces
physics incompatible with the stability of atoms. 

Breit performed a heuristic quantization of the Darwin interaction
energy, Eq. (\ref{eq:Darwin=000020interaction}), by inserting the
relativistic velocity operator from Eq. (\ref{eq:velocity=000020operator})
to give\cite{Breit_PhysRev1929}
\begin{equation}
\hat{g}\left(1,2\right)=\hat{g}^{C}\left(1,2\right)+\hat{g}^{B}\left(1,2\right);\quad\hat{g}^{B}\left(1,2\right)={\displaystyle -\frac{e^{2}}{4\pi\varepsilon_{0}}\left\{ \frac{c\boldsymbol{\alpha}_{1}\cdot c\boldsymbol{\alpha}_{2}}{2c^{2}r_{12}}+\frac{\left(c\boldsymbol{\alpha}_{1}\cdot\mathbf{r_{12}}\right)\left(c\boldsymbol{\alpha}_{2}\cdot\mathbf{r}_{12}\right)}{2c^{2}r_{12}^{3}}\right\} }.\label{eq:rel2int}
\end{equation}
 It should be emphasised, though, that since the interaction is truncated,
it is not Lorentz invariant. In practice, very often only the instantaneous
Coulomb two-electron interaction $\hat{g}^{C}$ , Eq. (\ref{eq:g^C}),
is included, giving the Dirac--Coulomb Hamiltonian.\cite{Saue_CPC2011}

Despite its relativistic origins, minimal electromagnetic coupling
is also used to introduce fields into the non-relativistic free-particle
Hamiltonian, giving
\begin{equation}
\hat{h}^{NR}=\frac{\hat{p}^{2}}{2m_{e}}+\frac{e}{2m_{e}}\left[\hat{\boldsymbol{p}}\cdot\boldsymbol{A}+\boldsymbol{A}\cdot\hat{\boldsymbol{p}}\right]+\frac{e^{2}A^{2}}{2m_{e}}-e\phi.
\end{equation}
Spin interactions are completely absent from this Hamiltonian, but
can be introduced by observing the identity
\begin{equation}
\left(\boldsymbol{\sigma}\cdot\hat{\boldsymbol{p}}\right)^{2}=\hat{p}^{2}I_{2},\label{eq:Dirac=000020identity}
\end{equation}
where $\boldsymbol{\sigma}$ are the Pauli spin matrices and $I_{2}$
is the $2\times2$ identity matrix. Performing minimal substitution
starting from the left-hand expression above one obtains
\begin{equation}
\hat{h}^{NR}=\frac{\hat{p}^{2}}{2m_{e}}+\frac{e}{2m_{e}}\left[\hat{\boldsymbol{p}}\cdot\boldsymbol{A}+\boldsymbol{A}\cdot\hat{\boldsymbol{p}}\right]+\frac{e^{2}A^{2}}{2m_{e}}+\frac{e}{2m_{e}}\left(\boldsymbol{\sigma}\cdot\boldsymbol{B}\right)-e\phi,\label{eq:NRham}
\end{equation}
with the appearance of the spin-Zeeman term. The identity Eq.(\ref{eq:Dirac=000020identity})
suggests that spin is somehow hidden in the non-relativistic Hamiltonian
and only manifests itself upon the introduction of a magnetic field.
This shows the economy of Nature's laws: we do not need to know if
a free particle has spin since it has nothing to interaction with.
This is in line with the observation that the free-particle Hamiltonian,
relativistic or not, does not contain information about the charge
of the particle, since this information is not needed in the absence
of fields.

We have seen that the Hamiltonian of Eq. (\ref{eq:NRham}) is based
a non-relativistic description of particles, but a relativistic coupling
of particles to fields. This is a perfectly reasonable pragmatic choice,
allowing the successful calculation of magnetic properties in a non-relativistic
framework. However, one may wonder what a truly non-relativistic coupling
of particles to fields would look like. What is the non-relativistic
limit of electrodynamics ? Formally, this amounts to taking the non-relativistic
limit of Maxwell's equations
\begin{equation}
\begin{array}{lclcclclc}
\boldsymbol{\nabla}\cdot\boldsymbol{E} & = & \rho/\varepsilon_{0} & (a) & ; & \boldsymbol{\nabla}\times\boldsymbol{E}+\partial_{t}\boldsymbol{B} & = & \boldsymbol{0} & (c)\\
\boldsymbol{\nabla}\cdot\boldsymbol{B} & = & 0 & (b) & ; & \boldsymbol{\nabla}\times\boldsymbol{B}-c^{-2}\partial_{t}\boldsymbol{E} & = & \frac{1}{\varepsilon_{0}c^{2}}\boldsymbol{j} & (d)
\end{array}.
\end{equation}
Taking the limit $c\rightarrow\infty$, we obtain
\begin{equation}
\begin{array}{lclcclclc}
\boldsymbol{\nabla}\cdot\boldsymbol{E} & = & \rho/\varepsilon_{0} & (a^{\prime}) & ; & \boldsymbol{\nabla}\times\boldsymbol{E} & = & \boldsymbol{0} & (c^{\prime})\\
\boldsymbol{\nabla}\cdot\boldsymbol{B} & = & 0 & (b^{\prime}) & ; & \boldsymbol{\nabla}\times\boldsymbol{B} & = & \boldsymbol{0} & (d^{\prime})
\end{array}.
\end{equation}
If we combine the observation that the magnetic field has zero divergence
\emph{and} zero curl with the boundary condition that the fields should
go to zero at infinite distance from the sources, we arrive at the
conclusion that the magnetic field is zero. Furthermore, all effects
of retardation vanish. Electrodynamics reduce to electrostatics, that
is, the instantaneous Coulomb interaction.\cite{Saue_AQC2005} As
such, Coulomb gauge is nice in that this non-relativistic limit is
embodied in the scalar potential, whereas all relativistic effects
are relegated to the vector potential. 

It should be added that taking the non-relativistic limit of Maxwell's
equations is complicated by the fact that the speed of light appears
in different positions according to the unit system chosen. In the
above equations, using full SI units, I conveniently eliminated the
magnetic constant $\mu_{0}$ using the relation $\mu_{0}\varepsilon_{0}=c^{-2}$.
However, in any unit system the vector potential appears with at least
an inverse power of the speed of light, as seen in Eqs. (\ref{eq:Lorentz=000020gauge=000020potentials})
and (\ref{eq:Coulomb=000020gauge=000020potentials}), hence in support
of the above conclusion. It appears that the Old Masters knew. Heisenberg
writes:\cite{Heisenberg1930}
\begin{quote}
\textsl{No attempt will be made to include relativistic effects, and
it is then logical to treat only electrostatic forces and to neglect
magnetic and retardational phenomena. }
\end{quote}
And in a much cited quote Dirac writes:\cite{Dirac_PRSA1929}
\begin{quote}
\textsl{The general theory of quantum mechanics is now almost complete,
the imperfections that still remain being in connection with the exact
fitting in of the theory with relativity ideas. These give rise to
difficulties only when high-speed particles are involved, and are
therefore of no importance in the consideration of atomic and molecular
structure and ordinary chemical reactions, in which it is, indeed,
usually sufficiently accurate }\textbf{\textsl{if one neglects relativity
variation of mass with velocity and assumes only Coulomb forces between
the various electrons and atomic nuclei}}\textsl{. The underlying
physical laws necessary for the mathematical theory of a large part
of physics and the whole of chemistry are thus completely known, and
the difficulty is only that the exact application of these laws leads
to equations much too complicated to be soluble. }
\end{quote}
The above emphasis is by me. In passing we note that the quote also
shows that Dirac did not believe that relativistic effects would be
of importance in chemistry since they arise in the core region.\cite{Kutzelnigg_Dirac}

Other authors have suggested that magnetic fields can exist in the
non-relativistic limit.\cite{Levy-Leblond1967,LeBellac_NC1973,Kutzelnigg_ch2002,Kutzelnigg_NMR_EPR_2004}
Unfortunately, Nature can not help us to resolve this issue since
we only have a relativistic universe available. However, we can at
least insist on consistency. Suppose that we introduce an external
uniform magnetic field through the vector potential
\begin{equation}
\boldsymbol{A}_{\boldsymbol{B}}=\frac{1}{2}\left(\boldsymbol{B}\times\boldsymbol{r}\right).\label{eq:A_B}
\end{equation}
 The non-relativistic Hamiltonian, Eq. (\ref{eq:NRham}), then becomes
\begin{equation}
\hat{h}^{NR}=\frac{\hat{p}^{2}}{2m_{e}}+\frac{e}{2m_{e}}\left(\hat{\boldsymbol{\ell}}\cdot\boldsymbol{B}\right)+\frac{e^{2}}{2m_{e}}\left[B^{2}r^{2}-\left(\boldsymbol{B}\cdot\boldsymbol{r}\right)\left(\boldsymbol{r}\cdot\boldsymbol{B}\right)\right]+\frac{e}{2m_{e}}\left(\boldsymbol{\sigma}\cdot\boldsymbol{B}\right),
\end{equation}
allowing the calculation of the full Zeeman interaction. The situation
is more complicated in the case of the vector potential associated
with a point-like nuclear magnetic dipole $\boldsymbol{m}_{K}$
\begin{equation}
\boldsymbol{A}_{K}=\frac{1}{4\pi\varepsilon_{0}c^{2}}\frac{\boldsymbol{m}_{K}\times\boldsymbol{r}_{K}}{r_{K}^{2}},\label{eq:A_K}
\end{equation}
because now the square of the speed of light appears in the denominator.
The usual approach is to keep the finite, relativistic speed of light.
This allows for instance to calculate NMR parameters in a non-relativistic
framework, but it does point to an inconsistency. In fact, a major
difference between the vector potentials of Eq. (\ref{eq:A_B}) and
Eq.(\ref{eq:A_K}) is that the latter contains the \emph{source} of
the magnetic field, the former not. If we instead of introducing an
\emph{external} magnetic field through Eq. (\ref{eq:A_B}) were to
include its source, e.g. a current loop, into our system, then no
magnetic interaction would occur at all. A prime example of this is
spin-orbit interaction: In a molecule an electron in its rest frame
sees charges, the nuclei or other electrons, in relative motion. These
therefore generate magnetic fields which interacts with the spin of
the electron. Upon setting the speed of light to some large value
in an actual relativistic calculation, the spin-orbit splittings vanish,
demonstrating that magnetic induction is a relativistic effect. In
the Pauli Hamiltonian the spin-orbit operator associated with some
nucleus of charge $Z$ is given by
\begin{equation}
\hat{h}_{SO}=\frac{Ze^{2}}{4\pi\varepsilon_{0}m_{2}^{2}c^{2}r^{3}}\boldsymbol{\sigma}\cdot\mathbf{\boldsymbol{\ell}}.\label{eq:h_SO}
\end{equation}
The operator form often leads to the interpretation that spin-orbit
interaction is the coupling of the spin and orbital angular momentum
of the electron. However, this is a misconception. As stated above,
spin-orbit interaction is magnetic induction, whereas spin-orbit coupling
is a \emph{consequence} of this interaction. In the above operator
the orbital angular momentum operator describes the relative motion
of electron and nucleus. Interestingly, this operator form ($\sim\boldsymbol{\sigma}\cdot\mathbf{\boldsymbol{\ell}}$)
is absent in the Dirac Hamiltonian, Eq. (\ref{eq:Rham}), so how is
spin-orbit interaction generated ? Here comes the beauty of relativistic
theory: The Hamiltonian is formulated in the frame of static nuclei
where only their electrostatic potential appears, but since the operator
obeys the Lorentz transformation, it captures the correct physics
in any frame, including the induced magnetic field felt by the electron
in its rest frame.

Finally, what about spin ? Is that a relativistic effect as well ?
We have seen that non-relativistic theory can perfectly well accomodate
spin (as can also be inferred by group-theoretical means \cite{Levy-Leblond1967}).
On the other hand, if magnetic fields vanish in the non-relativistic
limit, then spin has nothing to interact with, and then, in the spirit
of the economy of Nature's laws or using Occam's razor, we should
discard spin. However, the situation is not so straightforward, because
loosing spin implies loosing spin statistics and the distinction between
bosons and fermions, between symmetry and anti-symmetry. In this respect
it should be noted that several proofs of the spin-statistics theorem
are reported in the literature, some of them in a non-relativistic
framework, whereas most stress the crucial role of relativity in the
connection between spin and statistics.\cite{Curceanu_AJP2012} 

\section{More physics ?}

\subsection{Parity violation}

In the previous two sections we have seen that quantum mechanics was
needed in order to understand the periodic table of the chemical elements.
A proper description of the elements in the lower part of the table
even calls for the inclusion of the special theory of relativity.
It is therefore natural to ask if chemistry needs more physics.\cite{Pyykko_CR2012}
This question becomes even more pressing at a time where relativistic
quantum chemistry has reached a level of maturity where it may reach
the accuracy of non-relativistic molecular calculations. 

An interesting development is that fundamental physics, possibly beyond
the Standard Model, can be probed by tabletop spectroscopic experiments
instead of using gigantic particle accelerators like the Large Hadron
Collider, \cite{Safronova_RMP18} but this requires input from atomic
or molecular calculations. An intriguing example is provided by parity
violation in chiral molecules. Quantum chemistry is normally restricted
to electromagnetic interactions, but can be extended to include the
parity-violating component of the weak interaction, forming electroweak
quantum chemistry.\cite{Bakasov_JCP1998} In this framework a minute
energy difference is predicted between enantiomers of chiral molecules.
Such a mechanism has been proposed to explain the origin of biochirality,
that is, the observation that life is completely dominated by D-sugars
and L-amino acids, and not the opposite enantiomeric forms.\cite{Yamagata:1966:jtheobio}
There are alternative theories, though, and it is not clear how to
experimentally verify which one is at the origin of biochirality.
\cite{Martinez2022,Sallembien_CSR2022}

Starting at the level of quarks, an effective Hamiltonian for the
weak interaction between electrons and nuclei can be constructed.\cite{Berger_ch4_2004,bast:pncana_PCCP2011}
Here I only report the nuclear spin-independent part associated with
nucleus $A$
\begin{equation}
\hat{h}_{PV}^{A}=\frac{G_{F}}{2\sqrt{2}}Q_{W}^{A}\gamma^{5}\rho^{A}\left(\boldsymbol{r}\right).
\end{equation}
where $\rho^{A}$ is the nuclear charge distribution normalized to
unity, $\gamma^{5}$ is the fifth gamma matrix and 
\begin{equation}
Q_{W}^{A}=Z_{A}\left(1-4\sin^{2}\theta_{W}\right)-N_{A}
\end{equation}
is the weak nuclear charge, expressed in terms of the number of protons
($Z_{A}$) and neutrons ($N_{A}$), as well as the weak mixing angle
($\sin^{2}\theta_{W}=$0.22290(30)\cite{Tiesinga_RevModPhys.93.025010}).
The value of the Fermi coupling constant\cite{Tiesinga_RevModPhys.93.025010}
\begin{equation}
G_{F}=2.22252\times10^{-14}\,\text{E}_{h}a_{0}^{3}
\end{equation}
shows that the interaction is indeed weak and operates at an energy
scale several orders of magnitudes below the chemical one. The only
reason why we can hope to observe these extremely small effects in
atomic and molecular systems is that they break inversion. Various
experiments have been proposed to detect parity violation in chiral
molecules, but no successful experiment has been reported so far.\cite{Cournol_QE2019,Berger_Wire2019}

\subsection{Quantum Electrodynamics}

An ideal case for quantum chemistry is a molecule at 0K in the vacuum,
but is the vacuum really empty ?\cite{Greiner_PuZ1978} It was shown
by Ebbesen and co-workers that placing a molecule in a cavity changes
its reactivity.\cite{Hutchison_AngChem2012} The origin of this surprising
observation is the coupling of the molecule to the zero-point vibrations
of the quantized electromagnetic field. This has led to the development
of the exciting field of polaritonic chemistry.\cite{Ebbesen_CR2023,Ruggenthaler_CR2023}
Quantization of the electromagnetic field is also a prerequisite for
the theoretical study of spontaneous emission,\cite{sakurai:qed,craig:mqed}
without the introduction of phenomenological damping parameters.\cite{Norman_JCP2001}
\begin{center}
\begin{figure}
\begin{centering}
\includegraphics[width=0.5\textwidth]{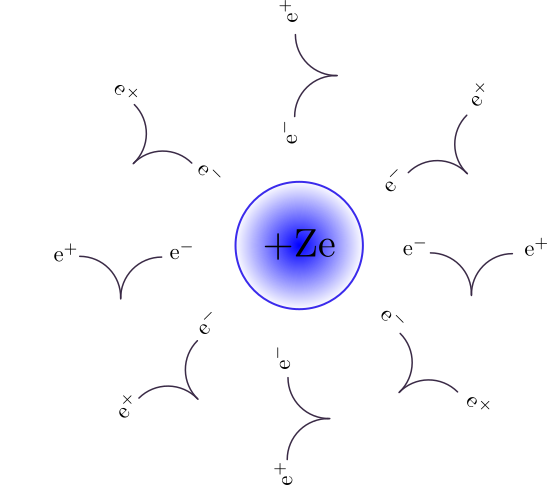}
\par\end{centering}
\caption{A pictorial representation of vacuum polarization.}\label{fig:VP}

\end{figure}
\par\end{center}

In the present context we shall rather focus on effects giving rise
to the Lamb shift: In the non-relativistic hydrogen-like atom the
$2s$ and $2p$ orbitals are degenerate.\cite{Bander_RevModPhys.38.330}
Switching to the Dirac equation the $2p$ shell is split by spin-orbit
interaction into $2p_{1/2}$ and $2p_{3/2}$ components, but the former
remains degenerate with the $s$-orbital, now denoted $2s_{1/2}$.
However, in 1947 Lamb and Retherford showed experimentally that this
degeneracy is broken as well, requiring physics beyond Dirac theory
for its explanation.\cite{Lamb_PhysRev.72.241} The missing physics
is provided by quantum electrodynamics (QED). From \emph{special relativity
}we learn of the equivalence of energy and mass, notably the conversion
of the energy of a photon to an electron-positron pair\cite{Hubbel_RPC2006}.
This requires an inordinate amount of energy, at least $2m_{e}c^{2}$,
corresponding to about 1 MeV, which is far beyond any reasonable energy
scale in chemistry. However, from the time--energy uncertainty principle
of \emph{quantum mechanics}, we learn that this energy can be ``borrowed''
for a limited amount of time. The vacuum is therefore far from empty,
passive space, rather a dynamic medium seething with virtual electron-positron
pairs. If we place a charge in space, it will be surrounded by such
pairs, as sketched in Figure \ref{fig:VP}. This \emph{vacuum polarization}
can not be disentangled from the ``bare'' charge of the particle
and therefore contributes to the \emph{observed charge}. In the same
manner an electron drags along its electromagnetic field which contributes
to its \emph{observed mass}; this effect is known as the \emph{self-energy
of the electron}. In hydrogen the Lamb shift is a mere 4 $\mu$eV,
but in hydrogen-like uranium it has grown to 76 eV.\cite{Yerokhin_JPCRD2015}
It is therefore justified to ask if QED effects can play a role in
chemistry.

With the consideration of QED, we move into the realm of quantum field
theory and new perspectives, nicely summarized by Frank Wilczek\cite{Wilczek_FR2006}
\begin{quote}
\textsl{In quantum field theory, the primary elements of reality are
not individual particles, but underlying fields. Thus, for example,
all electrons are but excitations of an underlying field, naturally
called the electron field, which fills all space and time. This formulation
explains why all electrons everywhere and for all time have exactly
the same properties ...}
\end{quote}
Furthermore, in QED, the electromagnetic field, now quantized, assumes
a more independent role.\cite{Wald_EM} In classical electrodynamics,
solving Maxwell's equations in the absence of sources gives electromagnetic
waves as solution. However, we can think of this of an idealised situation,
with the sources generating the fields being sufficiently remote so
that we may ignore them. Already in this classical framework, the
electromagnetic field can be described as a collection of harmonic
oscillators, but upon quantization the electromagnetic field acquires
zero-point energy and the vacuum fluctuates, also in the absence of
matter.\cite{Moss,Milonni_Qvac}

The Hamiltonian of QED describes the free electron and free photon
fields and their interaction\cite{Dyson_PhysRev.75.486}
\begin{equation}
\hat{H}=\hat{H}_{0}+\hat{H}_{\text{int}};\quad\hat{H}_{0}=\hat{H}_{0}^{\text{electron}}+\hat{H}_{0}^{\text{photon}}\label{eq:QED=000020Hamiltonian}
\end{equation}
In bound-state QED, the free-electron Dirac Hamiltonian is replaced
by the Dirac Hamiltonian in a classical background field, such as
the electrostatic field of clamped nuclei. This is known as the Furry
picture.\cite{Furry_PhysRev.81.115} Inserting such a Hamiltonian
into a time-dependent wave equation is problematic in view of Lorentz
invariance.\cite{Dyson_PhysRev.75.486} If we follow our system from
initial to final time within a given spatial radius, then we are effectively
focusing on a 4-dimensional cylinder in spacetime. However, under
a Lorentz transformation, this cylinder will deform. The flat ends
of the original cylinder, corresponding to the chosen time slices,
are examples of so-called space-like surfaces: a signal connecting
any two points of such a surface would have to go faster than the
speed of light, which is not allowed. The proper wave equation reads
\begin{equation}
i\hbar c\frac{\delta\Psi\left[\sigma\right]}{\delta\sigma\left(x\right)}=\hat{H}\left(x\right)\Psi\left[\sigma\right],
\end{equation}
where $x$ is a point in space-time and $\sigma$ a space-like surface
containing this spacetime point. This is the Schwinger--Tomonaga
equation where the action of the Hamiltonian on the wave function
is connected to the change in the wave function upon a modification
of the space-like surface, a special case being the time-derivative.

QED calculations are, however, conventionally carried out using perturbation
theory within S-matrix theory. We may introduce an operator that connects
the wave function at two different instances of time
\begin{equation}
|\Psi_{I}\left(t\right)\rangle=\hat{U}\left(t,t_{0}\right)|\Psi_{I}\left(t_{0}\right)\rangle;\quad t>t_{0},
\end{equation}
and satisfies the operator equation 
\begin{equation}
i\hbar\partial_{t}\hat{U}\left(t,t_{0}\right)=\hat{V}_{I}\left(t\right)\hat{U}\left(t,t_{0}\right);\quad\hat{V}_{I}=e^{+i\hat{H}_{0}t/\hbar}\hat{H}_{\text{int}}e^{-i\hat{H}_{0}t/\hbar}.
\end{equation}
The subscript $'I'$ indicates that we are in the interaction representation.
A formal solution reads
\begin{equation}
\hat{U}\left(t,t_{0}\right)=1+\frac{1}{i\hbar}\int_{t_{0}}^{t}dt^{\prime}\hat{V}_{I}\left(t^{\prime}\right)\hat{U}\left(t^{\prime},t_{0}\right).
\end{equation}
The equation can be iterated to give{\small
\begin{eqnarray}
\hat{U}\left(t,t_{0}\right) & = & 1+\frac{1}{i\hbar}\int_{t_{0}}^{t}dt_{1}\hat{V}_{I}\left(t_{1}\right)+\frac{1}{\left(i\hbar\right)^{2}}\int_{t_{0}}^{t}dt_{1}\int_{t_{0}}^{t_{1}}dt_{2}\hat{V}_{I}\left(t_{1}\right)\hat{V}_{I}\left(t_{2}\right)+\ldots\quad t_{1}\le t\\
 & \rightarrow & \sum_{n=0}^{\infty}\frac{1}{n!}\frac{1}{\left(i\hbar\right)^{n}}\int_{t_{0}}^{t}dt_{1}\int_{t_{0}}^{t}dt_{2}\ldots\int_{t_{0}}^{t}dt_{n}\hat{T}\left[\hat{V}_{I}\left(t_{1}\right)\hat{V}_{I}\left(t_{2}\right)\ldots\hat{V}_{I}\left(t_{n}\right)\right]e^{\epsilon\left(\left|t_{1}\right|+\left|t_{2}\right|+\ldots\left|t_{n}\right|\right)/\hbar},\label{eq:U-expansion}
\end{eqnarray}
}where in the final step \emph{time-ordering} is introduced so that
all integrals have not only the same lower limit, but also the same
upper limit. Also, an infinitesimal damping factor $\epsilon>0$ is
introduced to assure that the interaction is turned off at initial
and final times. A special case of the time-evolution operator is
the scattering or S-matrix
\begin{equation}
\hat{S}=\hat{U}\left(+\infty,-\infty\right),
\end{equation}
where initial and final times are the infinite past and the infinite
future, respectively, in order to assure Lorentz invariance. 

The interaction operator of QED reads
\begin{equation}
\hat{V}_{I}\left(t\right)=\int\hat{j}^{\mu}\left(x\right)\hat{A}_{\mu}\left(x\right)d^{3}\boldsymbol{x}\quad x=\left(ct,\boldsymbol{x}\right),
\end{equation}
 and connects the photon operator $\hat{A}_{\mu}$ with the electronic
4-current
\begin{equation}
\hat{j}^{\mu}\left(x\right)=-ec\hat{\psi}^{\dagger}\left(x\right)\alpha^{\mu}\hat{\psi}\left(x\right);\quad\alpha^{\mu}=\left(I_{4},\boldsymbol{\alpha}\right).
\end{equation}
The 4-current contains the electronic field operator $\hat{\psi}\left(x\right)$
and its adjoint $\hat{\psi}^{\dagger}\left(x\right)$ which in turn
may be expanded in some suitable orbital set
\begin{equation}
\hat{\psi}^{\dagger}\left(x\right)=\sum_{p}\varphi_{p}^{\dagger}\left(x\right)\hat{a}_{p}^{\dagger};\quad\hat{\psi}\left(x\right)=\sum_{p}\varphi_{p}\left(x\right)\hat{a}_{p}.
\end{equation}
The expansion coefficients $\hat{a}_{p}^{\dagger}$ and $\hat{a}_{p}$
are operators as well, namely the creation and annihilation operators,
that are well-known in the second quantization formulation of quantum
chemistry.\cite{helgaker:bok} Further connections to quantum chemistry,
notably for practitioners of coupled cluster theory,\cite{Crawford_RCC2000}
are made by noting that the time-ordered operator strings $\hat{T}\left[\ldots\right]$
appearing in (\ref{eq:U-expansion}) may be expanded, using Wick's
theorem,\cite{Wick_PhysRev1950} into of normal-ordered strings, usually
represented in terms of diagrams, in this case the iconic Feynman
diagrams. A challenge, though, is the appearance of divergences, associated
with loops in such diagrams. In modern QED such divergencies are tamed
in two steps: In a first step divergent integrals are made finite
by \emph{regularisation}, e.g. by the introduction of some cutoff
parameter. The expressions now depend on the values of the chosen
regularisation parameters. In the second step of \emph{renormalisation}
these parameters are absorbed into known observables, such as the
charge and mass of the electron.

The alert reader may note that the Hamiltonian of QED, Eq. (\ref{eq:QED=000020Hamiltonian}),
does not fit into the generic form of the electronic Hamiltonian given
by Eq.(\ref{eq:electronic=000020Hamiltonian}). In particular, there
is no two-electron interaction term. This is a necessary ingredient
of quantum field theory, as pointed out already by Dirac\cite{Dirac_RSPA1932}
\begin{quote}
\textsl{Classical electrodynamics ...teaches us that the idea of an
interaction energy between particles is only an approximation and
should be replaced by the idea of each particles emitting waves, which
travel outward with a finite velocity and influence the other particles
in passing over them.}
\end{quote}
Dirac next set out to demonstrate his point by introducing a one-dimensional
model in which two relativistic charges are coupled to the same quantized
field. By a perturbational treatment of the resulting time-dependent
wave equation, he could show that the lowest-order contribution in
the charges was equivalent to a wave equation featuring a Hamiltonian
with an explicit two-electron interaction. Indeed, if we consider
no-photon QED, that is, where there is no change in photon number
between the initial and final state, a lowest-order diagram is single-photon
exchange (cf. Figure \ref{fig:SP}), depicting the interaction of
two bound electrons (double blue lines) mediated by a photon (orange
curly line), formally represented by a photon propagator $D_{F}$,
as seen in the explicit S-matrix expression 
\begin{equation}
\hat{{\cal S}}_{SP}^{[2]}=-\frac{1}{2\hbar^{2}c^{2}}\int d^{4}x_{1}\int d^{4}x_{2}:\hat{j}^{\mu}\left(x_{1}\right)D_{F}\left(x_{1}-x_{2}\right)\hat{j}_{\mu}\left(x_{2}\right):e^{-\frac{\epsilon}{\hbar}\left(\left|t_{1}\right|+\left|t_{2}\right|\right)},
\end{equation}
where the beginning and the end of normal ordering are indicated by
colons. The black dots (vertices) in the diagram represent the two
spacetime points $x_{1}$ and $x_{2}$. In passing it may be interesting
to note that in his simple model, Dirac obtained an \emph{attractive}
electron-electron interaction, which we today know is a crucial feature
of superconductivity (Cooper pairs). In the regular manuscript Dirac
refers to this as a `mistake in sign', but, in a note added in proof,
Dirac cites Heisenberg stating that the result is correct since this
is what is expected for a one-dimensional, necessarily longitudinal
field. 
\begin{center}
\begin{figure}
\begin{centering}
\subfloat[Single-photon exchange\label{fig:SP}]{\begin{centering}
\includegraphics[totalheight=0.15\textheight]{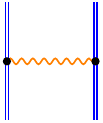}
\par\end{centering}
}\hfill{}\subfloat[Vacuum polarization\label{fig:VP-1}]{\begin{centering}
\includegraphics[totalheight=0.15\textheight]{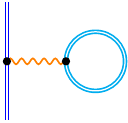}
\par\end{centering}
}\hfill{}\subfloat[Electron self-energy\label{fig:SE}]{\begin{centering}
\includegraphics[totalheight=0.15\textheight]{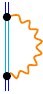}
\par\end{centering}
}
\par\end{centering}
\caption{Leading-order $\left(e^{2}\right)$ S-matrix contributions in the
no-photon case.}
\end{figure}
\par\end{center}

To the same order ($e^{2}$) appears the two QED-effects already mentioned.
In the vacuum polarization (VP) diagram (cf. Figure \ref{fig:VP-1})
one electron 4-current is replaced by a vacuum bubble, a virtual electron-positron
pair. Formally, it is represented by a Feynman electron propagator
(cyan double line) beginning and ending at the same space-time point,
as seen in the the explicit S-matrix expression
\begin{equation}
\hat{{\cal S}}_{VP}^{\left(2\right)}=\frac{1}{\hbar^{2}c^{2}}\int d^{4}x_{1}\int d^{4}x_{2}:\hat{j}^{\mu}\left(x_{1}\right):D^{F}\left(x_{1}-x_{2}\right)\text{tr}\left[-ec\beta\alpha_{\mu}S^{F}\left(x_{2},x_{2}\right)\right]e^{-\frac{\epsilon}{\hbar}\left(\left|t_{1}\right|+\left|t_{2}\right|\right)}.
\end{equation}
 Finally, there is a diagram representing the electron self-energy
(SE;cf. Figure \ref{fig:SE}). Here, we literally see an electron
emitting a photon at spacetime point $x_{2}$ and then reabsorbing
it at spacetime point $x_{1}$. Inbetween we see the (orange) curly
line of the photon propagator and the (cyan) double line of the electron
propagator. The S-matrix expression reads
\begin{equation}
\hat{{\cal S}}_{SE}^{\left(2\right)}=-\frac{e^{2}}{\hbar^{2}}\int d^{4}x_{1}\int d^{4}x_{2}e^{-\frac{\epsilon}{\hbar}\left(\left|t_{1}\right|+\left|t_{2}\right|\right)}:\hat{\psi}^{\dagger}\left(x_{1}\right)\alpha^{\mu}S^{F}\left(x_{1},x_{2}\right)\beta\alpha_{\mu}\hat{\psi}\left(x_{2}\right):D^{F}\left(x_{1}-x_{2}\right).
\end{equation}
The energy shifts associated with these QED effects can be calculated
from expectation values of these S-matrix operators.\cite{Sunaga_JCP2022}
In practice these will be expressed in terms of two-electron integrals
and, in the language of Hartree--Fock, the VP and SE contributions
correspond to direct and exchange contributions, respectively.

Although the S-matrix formalism of QED can rival experiment in accuracy,
it can not be applied to a general molecule since it does not combine
well with electron correlation.\cite{Blundell_PhysRevA.48.2615} In
recent years, a pragmatic approach has been devised using effective
QED-potentials that are added to the usual electronic Hamiltonian.
For vacuum polarization, such potentials have been known since the
early work of Uehling.\cite{Uehling_PR1935} Due to its exchange-like
(delocal) nature, the design of an effective potential for electron
self-energy is more challenging, but such potentials are now available.\cite{pekka:QEDmodel,Flambaum_PhysRevA.72.052115,Shabaev_PhysRevA.88.012513}
To get some idea of QED effects in chemistry, let us first look at
the ionization potential and electron affinity of gold. In Table \ref{tab:Au_IPEAdiff}
I have extracted data from Ref. \cite{Pasteka_PhysRevLett2017}. Calculations
using the coupled-cluster singles-and-doubles (CCSD) model based on
the Dirac--Coulomb (DC) Hamiltonian is not enough to get good agreement
with experiment, whereas the perturbative inclusion of triples, giving
CCSD(T), the `gold standard of quantum chemistry' overshoots. Extending
the Coulomb two-electron interaction, Eq. (\ref{eq:rel2int}a) by
the Breit term, Eq. (\ref{eq:rel2int}b) brings the calculated value
closer to experiment, but the crucial contributions are provided by
the inclusion of higher-order excitations to coupled-cluster as well
as the inclusion of QED-effects through effective potentials. These
two latter contributions are of the same order of magnitude. It suggests
that if one want to do relativistic coupled-cluster calculations with
an accuracy better than CCSD(T) one should also include QED effects. 
\begin{center}
\begin{table}
\begin{centering}
\begin{tabular}{l|cc|cc}
 & \textsf{IP} &  & \textsf{EA} & \tabularnewline
\hline 
\textsf{DC-CCSD} & \textsf{9.1164} &  & \textsf{2.1070} & \tabularnewline
\textsf{DC-CCSD(T)} & \textsf{9.2938} & \textsf{+0.1774} & \textsf{2.3457} & \textsf{+0.2387}\tabularnewline
\textsf{DC-CCSDTQP} & \textsf{9.2701} & \textsf{-0.0237} & \textsf{2.3278} & \textsf{-0.0179}\tabularnewline
\textsf{Breit} & \textsf{9.2546} & \textsf{-0.0155} & \textsf{2.3188} & \textsf{-0.0090}\tabularnewline
\textsf{QED} & \textsf{9.2288} & \textsf{-0.0258} & \textsf{2.3072} & \textsf{-0.0116}\tabularnewline
\hline 
\textsf{Experiment\cite{Brown_JOSA1978,Andersen_JPCRD1999}} & \textsf{9.2256} &  & \textsf{2.3086} & \tabularnewline
\end{tabular}
\par\end{centering}
\caption{Ionization potential and electron affinity of gold.\cite{Pasteka_PhysRevLett2017}}\label{tab:Au_IPEAdiff}

\end{table}
\par\end{center}

Another perspective is provided by Table \ref{tab:Au_IPEAcont}. Here
I compare relativistic and QED effects the ionization potential and
electron affinity of gold. We first see, in Table \ref{tab:Au_rel},that
the relativistic effects are dramatic, in particular for the electron
affinity, where they reach almost 80\%. Clearly, relativity has a
huge impact on the chemistry of a heavy element such as gold. The
effect of QED is far more modest, leading to a slight reduction of
the values. Results such as these led Pyykkö to propose a simple rule
of thumb: \emph{QED effects reduce relativistic effects by about 1\%.}\cite{Pyykk0_PhysRevA.57.R689}
\begin{center}
\begin{table}
\begin{centering}
\subfloat[Relativistic effects.\cite{Fossgaard_JCP2003}\label{tab:Au_rel}]{\selectlanguage{english}%
\begin{centering}
\begin{tabular}{l|cc}
\foreignlanguage{british}{} & \foreignlanguage{british}{IP/eV} & \foreignlanguage{british}{EA/eV}\tabularnewline
\hline 
\foreignlanguage{british}{NR} & \foreignlanguage{british}{7.064} & \foreignlanguage{british}{1.287}\tabularnewline
\foreignlanguage{british}{+R} & \foreignlanguage{british}{9.195} & \foreignlanguage{british}{2.301}\tabularnewline
\hline 
\foreignlanguage{british}{$\Delta_{rel}$} & \foreignlanguage{british}{30.2\%} & \foreignlanguage{british}{78.8\%}\tabularnewline
\end{tabular}
\par\end{centering}
\selectlanguage{british}%

}\quad{}\subfloat[QED-effects.\cite{Pasteka_PhysRevLett2017}\label{tab:Au_qed}]{\selectlanguage{english}%
\begin{centering}
\begin{tabular}{l|cc}
\foreignlanguage{british}{} & \foreignlanguage{british}{IP/eV} & \foreignlanguage{british}{EA/eV}\tabularnewline
\hline 
\foreignlanguage{british}{R} & \foreignlanguage{british}{9.2546} & \foreignlanguage{british}{2.3188}\tabularnewline
\foreignlanguage{british}{+QED} & \foreignlanguage{british}{9.2288} & \foreignlanguage{british}{2.3072}\tabularnewline
\hline 
\foreignlanguage{british}{$\Delta_{QED}$} & \foreignlanguage{british}{-0.28\%} & \foreignlanguage{british}{-0.50\%}\tabularnewline
\foreignlanguage{british}{{\small$\Delta_{QED/R}$}} & \foreignlanguage{british}{\textsf{-1.21\%}} & \foreignlanguage{british}{\textsf{-1.14}\%}\tabularnewline
\end{tabular}
\par\end{centering}
\selectlanguage{british}%
}
\par\end{centering}
\caption{Contributions beyond the non-relativistic (NR) value to the calculated
ionization potential and electron affinity of gold }\label{tab:Au_IPEAcont}

\end{table}
\par\end{center}

\begin{center}
\begin{figure}
\begin{centering}
\includegraphics[width=0.8\textwidth]{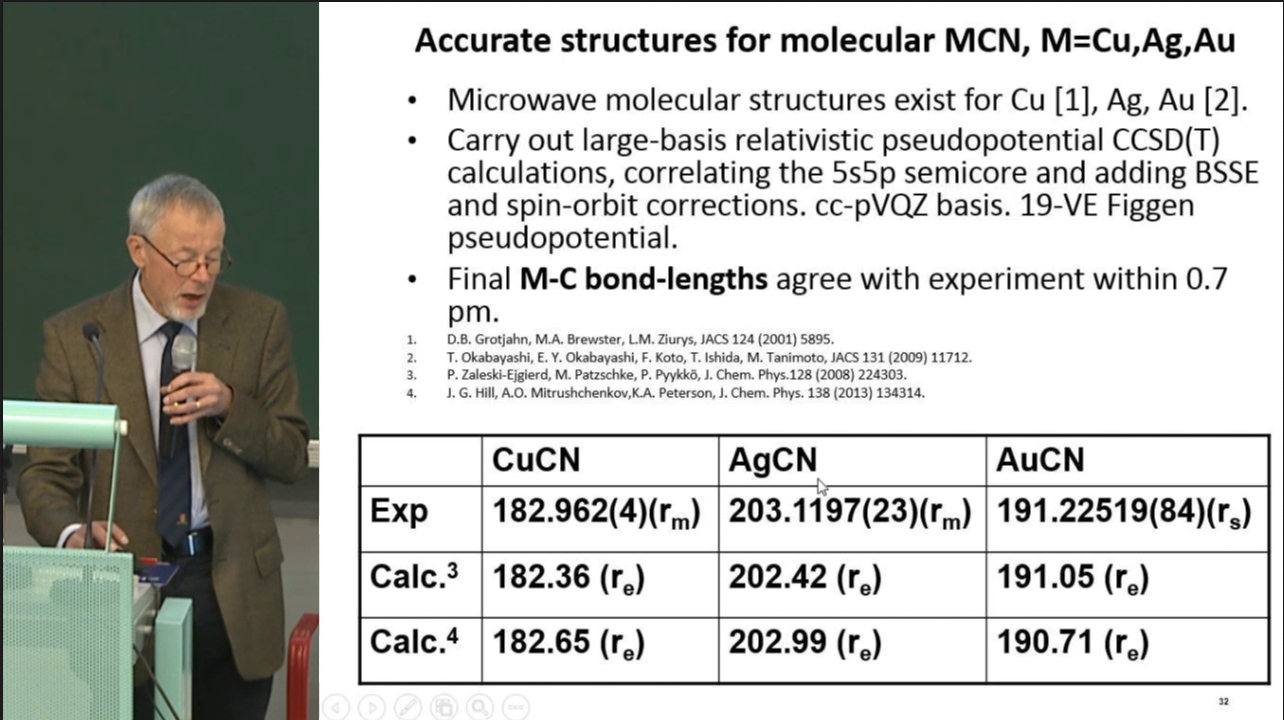}
\par\end{centering}
\caption{Pekka Pyykkö discussing a possible observable QED-effect during his
Davidson lecture in 2013 (with permission from P. Pyykkö).\cite{pyykkotalk}}\label{fig:Pyykk=0000F6-talk}
\end{figure}
\par\end{center}

What then about molecules ? In his 2013 Davidson lecture, delivered
for the University of North Texas, Pyykkö made the following observation
and conjecture (cf. Figure \ref{fig:Pyykk=0000F6-talk}): In 2008
Pyykkö and co-workers reported accurate calculations of the geometries
of the cyanides of the noble metals.\cite{Zaleski_JCP2008} In 2013
the group of Peterson presented even more accurate calculations on
the same molecules.\cite{Hill_JCP2013} For \ce{CuCN} and \ce{AgCN}
the calculated metal-carbon bond lengths moved closer towards the
values reported in a 2009 microwave experiment.\cite{Okabayashi_JACS2009},
but for \ce{AuCN} the discrepancy with experiment increased. Pyykkö
concluded that the latter result indicated missing physics and conjectured
that this was an instance of an observable QED-effect. Support for
this conjecture comes from his rule of thumb: A non-relativistic calculation
using density functional theory (DFT) gives an \ce{Au-C} bond length
of 218.54 pm, whereas the inclusion of relativity reduces it to 193.16
pm.\cite{Sunaga_JCP2022} Using his rule of thumb, the relativistic
bond length reduction of about 25 pm would indicate that QED would
increase the bond length by about 0.25 pm, which is very close to
the discrepancy between experiment and the best calculated value.
We recently implemented effective QED-potentials in the DIRAC code
for relativistic molecular calculations\cite{Sunaga_JCP2022,DIRACpaper2020}
(see also Ref. \cite{skripnikov_jcp154.201101}), and set out to investigate
this conjecture. The devil is in the details, though: As seen in Figure
\ref{fig:Pyykk=0000F6-talk} calculations reported equilibrium bond
distances $r_{e}$, whereas the microwave experiment for \ce{AuCN}
reported the substitution structure $r_{s}$ obtained by consideration
of the change of center-of-mass upon single isotope substitution.\cite{demaison2016equilibrium}
We therefore decided to calculate directly the rotational constants
$B_{0}$ extracted from the microwave spectra. The results are shown
in Table \ref{tab:AuCN_B0} and provides strong support in favor of
the conjecture made by Pyykkö.
\begin{center}
\begin{table}
\begin{centering}
{\footnotesize{}%
\begin{tabular}{lccc}
{\footnotesize$B_{0}$(MHz)} & {\footnotesize w/o QED} & {\footnotesize with QED} & {\footnotesize Exp.\cite{Okabayashi_JACS2009}}\tabularnewline
\hline 
{\footnotesize\ce{^197Au^12C^14N}} & {\footnotesize 3235.1} & {\footnotesize 3230.4} & {\footnotesize 3230.21115(18)}\tabularnewline
{\footnotesize\ce{^197Au^13C^14N}} & {\footnotesize 3182.1} & {\footnotesize 3177.5} & {\footnotesize 3177.20793(13)}\tabularnewline
{\footnotesize\ce{^197Au^12C^15N}} & {\footnotesize 3084.3} & {\footnotesize 3079.9} & {\footnotesize 3079.73540(12)}\tabularnewline
{\footnotesize$r_{s}$(\ce{Au-C})} & {\footnotesize 190.991} & {\footnotesize 191.184} & {\footnotesize 191.22519(84)}\tabularnewline
\end{tabular}}{\footnotesize\par}
\par\end{centering}
\caption{Calculated and experimental vibrational constants for isotopomers
of \ce{AuCN} as well as the resulting substitution bond length.\cite{Sunaga_JCP2022}}\label{tab:AuCN_B0}

\end{table}
\par\end{center}

Similar to the effects of relativity, QED effects are generated in
the vicinty of nuclei, but at much more reduced length scales, on
the order of a reduced Compton wave length $\lambdabar_{C}=\hbar/m_{e}c$.
So far, studies of QED effects have mostly focused on valence properties
such as spectroscopic constants or dipole moments, but in view of
the above observation, one may expect even stronger QED effects for
properties probing the electronic density in the vicinity of heavy
nuclei. A problem, though, is that such calculations may be outside
the domain of validity of the effective QED-potentials. We have therefore
initiated work towards a variational approach to QED (for other work
in the same direction, see for instance Refs. \cite{Liu_Lindgren_JCP2013,Toulouse_SP2021,Matyus_ACS2023,Flynn_PhysRevA.111.042810}).
We insist, though, on staying within the usual framework of relativistic
quantum chemistry, notably using the local Gaussian-type basis sets
introduced by Boys in 1951.\cite{BoysI} In this respect it may be
of interest to quote from a biography of Boys\cite{Hall_MP1996}
\begin{quote}
\textsl{Boys felt that his approach to quantum chemistry could be
extended to other areas of science. Indeed his first work with Shavitt
... was on the use of Gaussian-like expansions of the intermolecular
potential to compute virial coefficients. He even tried to bring Gaussians
into quantum field theory. He knew that the subject was plagued by
infinities which were sometimes the result of poor technique and sometimes
of poor modelling. His Gaussian-based treatment could get rid of some
of these but, when he talked to the local experts about his proposed
starting point, they indicated that one infinity remained in the operators
which he could not remove, so the entire treatment fell down. Nothing
came of it but he was not disheartened.}
\end{quote}
We have unfortunately not been able to learn more about this stifled
development, but do find that QED-calculations using Gaussian basis
sets can rival in accuracy with conventional ones based on analytic
solutions of the Dirac equation.\cite{Salman_PhysRevA.108.012808,Ferenc_PRA2025}

\section{Conclusion}

The reader, having worked through this mini-review, may ask: Where
is chemistry ? The purpose of the review has been to convey the physics
underlying the periodic table and then my fellow chemists can take
it from there. Accounts along these lines are already available,\cite{Pyykko_CR2012,Schwerdtfeger_NRC2020}
but I have tried to give a feeling for the underlying physics, to
add flesh to the mathematical skeleton, with a particular emphasis
on quantum effects. The physics provides the foundations, but then,
to quote Lehn\cite{Lehn_ANG2015}
\begin{quote}
\textsl{It is the task of chemistry to build the bridge between physics
and the general laws of the universe on one hand, and biology and
the emergence of life and thought on the other hand.}
\end{quote}
This is a creative process and the atoms are the colours of the palette
in the art of matter. To further cite Lehn \cite{Lehn_ANG2015}
\begin{quote}
\textsl{Chemistry continuously reinvents and recreates itself by recomposition
of the basic building blocks of matter, the elements displayed in
the Periodic Table, one of the greatest achievements of mankind$\ldots$and
the playground of chemistry !}
\end{quote}
Physicists have also expressed appreciation for our playground, notably
Feynman:\cite{Feynman:1963:FLP}
\begin{quote}
\textsl{If, in some cataclysm, all of scientific knowledge were to
be destroyed, and only one sentence passed on to the next generations
of creatures, what statement would contain the most information in
the fewest words? I believe it is the atomic hypothesis (or the atomic
fact, or whatever you wish to call it) that all things are made of
atoms---little particles that move around in perpetual motion, attracting
each other when they are a little distance apart, but repelling upon
being squeezed into one another. In that one sentence, you will see,
there is an enormous amount of information about the world, if just
a little imagination and thinking are applied.}
\end{quote}

\subsection*{Acknowledgements}

This project received funding from the European Research Council (ERC)
under the European Union’s Horizon 2020 research and innovation program
(Grant Agreement No. 101019907 HAMP-vQED). I thank Dávid Ferenc for
critical reading of the manuscript.

\selectlanguage{english}%
{} 

\printbibliography

\selectlanguage{british}%

\end{document}